\documentstyle[epsfig]{aipproc}

\def\beq{\begin{equation}}
\def\eeq{\end{equation}}
\def\bea{\begin{eqnarray}}
\def\eea{\end{eqnarray}}
\def\bq{\begin{quote}}
\def\eq{\end{quote}}

\def\NP{{\it Nucl.Phys.} }
\def\PL{{\it Phys.Lett.} }

\def\gappeq{\mathrel{\rlap {\raise.5ex\hbox{$>$}}
{\lower.5ex\hbox{$\sim$}}}}

\def\lappeq{\mathrel{\rlap{\raise.5ex\hbox{$<$}}
{\lower.5ex\hbox{$\sim$}}}}

\begin{document}
\title{Heavy-Ion Collisions and Black Holes in Anti-de-Sitter Space}

\author{John Ellis$^*$ }
\address{$^*$Theoretical Physics Division, CERN \\ CH -- 1211 Geneva 23}
\maketitle

\begin{center}
{\it Invited talk given at\\
`RHIC Physics and beyond - Kay Kay Gee Day'\\
Brookhaven National Laboratory, October 23rd, 1998}\\
{~~}\\
CERN-TH/98-391~~~~~~~~~~nucl-th/9902061\\
\end{center}

\begin{abstract}
Recent developments linking non-perturbative quantum gauge theories
in Minkowski space to classical gravity theories in anti-de-Sitter
space are reviewed at a simple level. It is suggested how these
spectacular advances may be extended to discuss the quark-gluon
phase transition in terms of black holes in anti-de-Sitter space,
with possible relevance to heavy-ion collisions.
\end{abstract}

\section*{Introduction}

{\tt Klaus Geiger was a totally special guy, and it was a privilege to
have him as a friend
as well as a valued collaborator.  Many of us at CERN have been deeply
distressed
by his passing away - including secretaries,  barmen and
playgoers as well as physicists.}  

In addition to our joint projects, Klaus and I used to enjoy
shooting the bull
about our other nutty ideas, not to mention string theory.  I am sure he would
enjoy hearing how the latest developments in this formal field might 
connect with his passion for RHIC physics.  Mapping QCD onto gravity in
anti-de-Sitter space may provide a new way of thinking about (even
calculating?) the quark-hadron transition. On the way to describing
this possibility in simple terms, I start by reviewing some ideas
about modelling this transition that Klaus and I incorporated in our
work together, and we shall see later how this may be related to
anti-de-Sitter black holes.

\section*{Effective Field Theory and the Quark-Hadron Transition}

The low-mass hadronic degrees of freedom, $\pi, K, \eta, ...$ and
their low-energy interactions are described by an effective chiral
Lagrangian~\cite{chilag}, that can be written in the form
\beq
{\cal L}_{\rm eff} = - \frac{1}{2} \partial_\mu \underline{\pi} \cdot \partial^\mu
\underline{\pi} + {\cal O}(\underline{\pi}^2(\partial \underline{\pi}^2))
+ ...~~.
\label{one}
\eeq
The quartic and higher-order terms may in 
principle be calculated by integrating out quarks,
or be related to the exchanges of more massive hadrons, such as the vector
mesons, and their form is being constrained by data~\cite{intout}.  The
effective Lagrangian
(\ref{one}) also has soliton (lump) solutions that have $I = J = 1/2$ and $B =
1$, that may be identified with  baryons~\cite{Skyrme}.  This picture may
provide useful
insight into the puzzle of the proton spin, which is purely orbital angular
momentum in this soliton approach, so that the quark spins make no net
contribution in the limit of light quark masses~\cite{BEK}, in qualitative
agreement with the data~\cite{nspin}.

At finite temperatures, the chiral order parameter $\langle 0|\bar q q |0
\rangle \rightarrow 0$ and chiral symmetry is restored~\cite{BG,GL}.  In
this limit, the
chiral soliton expands:  $R \rightarrow \infty$ and its mass vanishes:  $M
\rightarrow 0$.  We therefore interpret this chiral transition also as the
deconfinement transition for quarks~\cite{CEO}.  If only the pseudoscalar
degrees of
freedom in (\ref{one}) are taken into account, this chiral transition is second
order~\cite{BG,GL}.  However, the transition occurs at a temperature $T =
0(\Lambda_{QCD}),$
so other hadronic degrees of freedom should also be taken into
account~\cite{CEO}. 
Figure~1 shows how an order parameter $\chi$ related to the gluon condensate
$\langle 0|G_{\mu\nu} G^{\mu\nu}|0\rangle$ behaves as a function of temperature
when these are included.  At low temperature $T_0 < T_1$, the free energy is
minimized when $\chi \not= 0$ and the heavier hadrons with $m \propto \chi$ are
massive.  At $T_1$, the free energy is equal in this and the $\chi = 0$ phase
where the hadrons are all massless.  Below $T_2$, there is the possibility of a
mixed phase, depending on the history of the system.  Finally, when one heats
beyond $T_2$, the $\chi \not= 0$ confined vacuum ceases to exist, and the
theory must be in the unconfined phase.  This picture is reminiscent of a
liquid-gas transition~\cite{EGM}, and is analogous with phase transitions
of gravity in anti-de-Sitter space, as we shall see later.


\begin{figure}[htb]
\hspace*{10pt}
\centerline{\epsfig{file=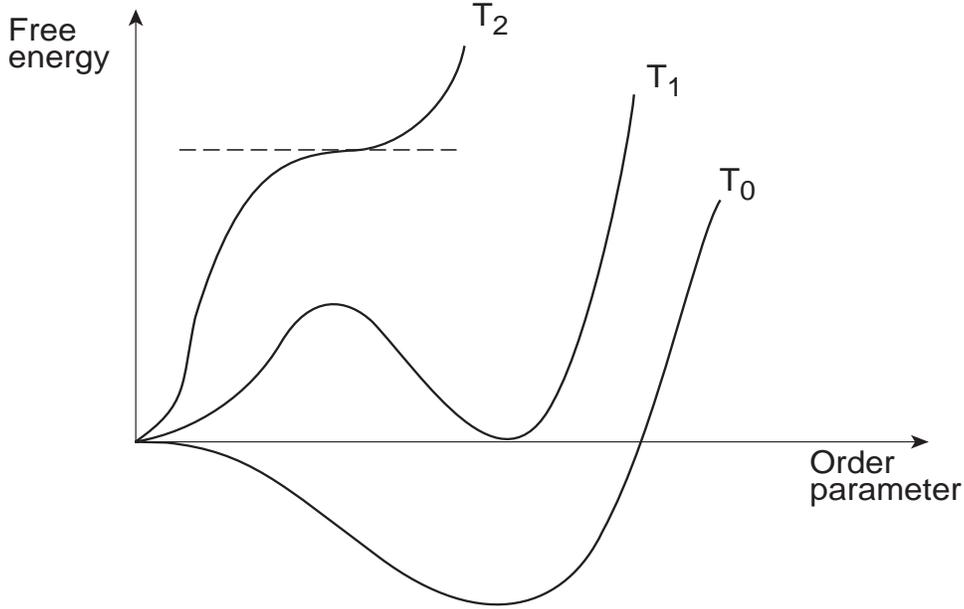,height=3.2in,width=5in}}
\vspace{0.5cm}
\caption[] { {\it Qualitative variation of the free energy with
temperature
in QCD~\cite{CEO}. At temperatures below $T_0$, there is no stable
state at the origin, the only stable state (with non-zero order parameter)
corresponds to confinement, and there
is a string tension. There is no stable
confined state at temperatures above $T_2$. Between $T_0$
and $T_2$, there is the possibility of a mixed phase. The confined
and unconfined phases have equal free energies at the temperature $T_1$,
which would be $T_c$ under adiabatic conditions.}}
\label{fig1}
\end{figure}

Klaus and I incorporated these ideas~\cite{EG} in his space-time
parton-shower Monte
Carlo~\cite{VNI}, to provide an afterburner for hadronization.  To see how
this works,
consider the process $e^+e^- \rightarrow Z^0 \rightarrow \bar q q$, where the
initial decay produces a pair of ``hot" off-shell partons as shown in Fig.~2a. 
These create a local ``hot spot" of the unconfined phase, which is surrounded
by the usual ``cold" hadronic vacuum.  During the subsequent parton-shower
development, the ``hotter" deeply-virtual partons decay into ``cooler" partons
closer to mass shell, as the ``hot spot" expands and cools.
 Since an isolated parton is impossible, whenever a parton becomes $\gappeq
 1$~fm away from its nearest neighbour, it should be confined and hadronize. 
 This process continues until all the partons ``cool", separate and hadronize,
 as seen in Fig.~2a for $e^+e^- \rightarrow Z^0 \rightarrow {\bar q} q$
and 
in Fig.~2b for $e^+e^- \rightarrow W^+ W^- \rightarrow {\bar q} q {\bar q}
q$.


\begin{figure}[htb]
\hspace*{30pt}
\centerline{\epsfig{file=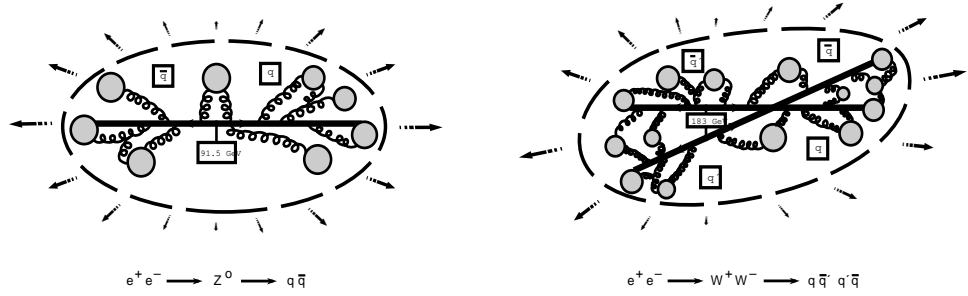,height=2in,width=5.3in}}
\vspace{0.5cm}
\caption[]{{\it
Schematics of the two $e^+e^-$ event types 
$e^+e^- \rightarrow Z^0 \rightarrow {\bar q} q$ and 
$e^+e^- \rightarrow W^+ W^- \rightarrow {\bar q} q {\bar q} q$.
The final-state hadron distribution in $Z^0$ events (left) is due to 
exclusively `endogamous' hadronization of the partonic offspring from 
the $q\bar{q}$ dijet~\cite{EG}, whereas in $W^+W^-$ events (right) there
is, in 
addition, the possibility of `exogamous' hadron production involving a 
mating of partons from the two different $W^+ \rightarrow q\bar{q}$ and
$W^- \rightarrow q\bar{q}$ dijets~\cite{EGWW}.}}
\label{fig2}
\end{figure}

 We modelled~\cite{EG} the parton-hadron conversion using the chiral
Lagrangian
 (\ref{one}) to describe the ``cool" phase, and describing the transition by
 analogy with the finite-temperature transition shown in Fig.~1, but
replacing temperature $T$ by the
 separation $L$ between a parton pair:  $T \leftrightarrow
\frac{1}{L}$~\cite{coffin}. We
 interpreted $L$ as the effective size of the local cooling region, and treated
 the hadronic degrees of freedom as if they were in a box of this size.  We
 then estimated heuristically the hadronization probability $P(L)$ by calculating the
 tunnelling through the barrier shown in Fig.~1 between the ``partonic" $\chi =
 0$ and hadronic $\chi \not= 0$ vacua.  This probability peaks at a
 characteristic separation $L \sim 0.6$ to 0.8~fm.  We have applied this
 picture to various physical conditions, including deep-inelastic
scattering~\cite{EGK}
 and $e^+e^- \rightarrow W^+W^- \rightarrow$ hadrons~\cite{EGWW}, as well
as to $e^+e^-
 \rightarrow Z^0 \rightarrow$ hadrons~\cite{EG}. An afterburner for
Bose-Einstein
 correlations~\cite{GEHW} that we explored together with Klaus is
described here by
 Ulrich~Heinz and Urs~Wiedemann~\cite{UHUW}.
 
In the rest of this talk, we explore a different direction, first
looking at more theoretical approaches to confinement, and seeing later
their possible relation to the ideas discussed in this section.

\section*{Confinement and Monopoles}
 
 A formal approach to confinement is based on the space-like Wilson loop
 integral
 \beq
 W(C) = e^{ie \int_cA\cdot dl}~,
 \label{two}
 \eeq
 and the time-like Polyakov loop integral
 \beq
 P(C') = e^{ie \int_{C'}A\cdot dl}~.
 \label{three}
 \eeq
 If we consider a curve $C'$ that has the form of a rectangle with time-like
 sides of duration $t$ separated by a distance $R$, in the confining phase we
 expect that
 \beq
 P(R,t) \sim e^{-\mu R t}~,
 \label{four}
 \eeq
 corresponding to an effective potential
 \beq
 V(R) \sim \mu R
 \label{five}
 \eeq
 This is believed to correspond to a flux tube in which the chromoelectric flux
 lines have been squeezed by a dual analogue of the Meissner effect in
 conventional superconductivity~\cite{monopoles}.  In the conventional
case,
there is a
 condensate of Cooper pairs $\langle e^-e^-\rangle$ that seeks to expel
 magnetic flux lines.  In the QCD case, there should be a condensate of
 chromomagnetic monopoles that seek to expel and squeeze chromoelectric flux.  Essential aspects of this picture have been
 demonstrated in $N = 2$ supersymmetric gauge theory, which has an exact
 non-perturbative solution~\cite{SW}.
 
 The chromomagnetic monopoles arise in the projection of a non-Abelian gauge
 theory onto its maximal Abelian subgroup, e.g., $SU(N) \rightarrow
U(1)^{N-1}$~\cite{monopoles}.
  It is believed that this Abelian subgroup may be solely responsible for
  confinement, and Polyakov has proposed a formal string
description of QCD in
  this confined phase~\cite{Pol}.  Assuming monopole condensation at a
characteristic QCD
  scale $\Lambda$, he has argued that low-energy physics is described by an
  effective action for a chiral gauge field, that includes an antisymmetric
  tensor field coupled to string:
  \beq
  A \rightarrow \int d^4 x B_{\mu\nu}(x) \int d^2 \sigma \epsilon_{ab}
  \partial_a X^\mu \partial_b X^\nu \delta^4 (X(\sigma) - x)
  \label{six}
  \eeq
  where $\sigma$ and the suffices $a,b$ denote world-sheet coordinates. 
  Relating this to the formalism of fundamental string theory, and deriving a
  full theory of QCD string, remain theoretical challenges.
  
  \section*{Introduction to String Theory}
  It is worth recalling that string models were born from old-fashioned
  strong-interaction theory.  The lowest-lying hadrons $p, n, \pi$ were found
  to be accompanied by many heavier hadronic states, whose maximum spins grew
  quadratically with their masses:
  \beq
  m^2 \propto J~,
  \label{seven}
  \eeq
  as would occur in a linearly-rising potential (\ref{five}).  Such a
potential
  would lead to an infinite set of unstable excited states.
  
  It was observed that such an infinite set of states would make possible a
  dual bootstrap picture in which the infinite set of direct-channel resonances
  could be resummed to yield an infinite set of cross-channel exchanges, 
which were associated with Regge trajectories.  A
  mathematical model scattering amplitude embodying these ideas was found
by
  Veneziano~\cite{Veneziano}.
  
  Shortly afterwards, it was shown that this amplitude could be derived from a
  string theory~\cite{string}, in which the highest-spin excited states
had
  \beq
  J \simeq \alpha'm^2~,
  \label{eight}
  \eeq
  where $\alpha' \sim$(1~GeV)$^{-2}$ is the string tension.  The hadrons could
  be interpreted in this picture as excited vibrational states of open strings
  with massless quarks attached to their ends.  Unitarity at higher orders
  required that these be supplemented by closed strings that were thought to be
  related to the Pomeron and to glueballs.
  
  Attention to this approach to the strong interactions was distracted by its
  failure to explain the point-like parton structures that had
emerged in high
  momentum-transfer processes.  Interest in the string approach to QCD lapsed
  with the discovery of QCD, which showed how this point-like behaviour could
  be understood in field theory through the property of asymptotic
freedom.
  
  However, around the same time it was proposed that string theory be
  reinterpreted as a quantum theory of gravity and all the other particle
  interactions~\cite{SS}.  This proposal arose from the observation that
the spectrum of
  closed string included a massless spin-2 particle that could be interpreted
  as the graviton, as well as massless spin-1 particles that could be
  interpreted as gauge bosons.  This reinterpretation apparently required
  rescaling the string tension:  $\alpha' \rightarrow {\cal
  O}\left(\frac{1}{m^2_P}
  \right)$, where $m_P \simeq 10^{19}$~GeV is the energy scale at which
  four-dimensional gravitational interactions become strong.
  
  String theories can be written in terms of two-dimensional field theories on
  the world sheet which describes the propagation of the string through time. 
  A scalar field on the world sheet of a closed string can be decomposed into
  left- and right-moving waves:
  \beq
  \phi(r+t) + \phi (r-t)~,
  \label{nine}
  \eeq
  each of which can be expanded in normal modes.
  Such a two-dimensional field theory is relatively easy to make finite at the
  loop level, thus offering the prospect of a finite quantum theory of gravity.
   This finiteness may be traced to the softening of divergence difficulties
   that occurs when point particles are replaced by extended structures, e.g.:
   \beq
   \int d^4k \frac{1}{k^2} \leftrightarrow \int_{1/\Lambda=R} 
   d^4x \frac{1}{x^6} \sim \frac{1}{R^2}~,
   \label{ten}
   \eeq
   where $R$ is the size of the extended object.  Strings are the simplest
   examples of this approach, but why not also consider higher-dimensional
   internal structures?  This is actually what happens when one confronts
   non-perturbative aspects of string theory~\cite{Polchinski}, as we
discuss later.
   
   In making a finite string theory, however, one must be careful to avoid
   anomalies at the quantum level.  In the case of a purely bosonic string with
   no additional world-sheet degrees of freedom, 
at the perturbative level this requires embedding it in
   26 space-time dimensions~\cite{Lovelace}.  If the string is made
supersymmetric, the 
   necessary number
   of dimensions is reduced to ten.  The complete quantum consistency
   of these theories was demonstrated by Green and Schwarz~\cite{GS},
triggering the    first string revolution.
   
   The original bosonic string suffered from the defects that it had no
   fermionic matter and that its vacuum was unstable, since it had a tachyonic
   scalar field.  Superstrings gained on both counts, since they contained
   fermions and had stable vacua with zero cosmological constant.  However,
   they were initially discarded because it was not seen how to incorporate
   parity violation.  This was first achieved with the ten-dimensional
   heterotic string~\cite{heterotic}, which retained the feature of a
stable zero-energy vacuum.
    The final step towards making realistic models came with the realization
    that one could construct four-dimensional heterotic string
models, either
    by compactifying the six surplus dimensions on a small manifold with
    characteristic size $\sim \frac{1}{m_P}$~\cite{quartet}, or by direct
construction in four
    dimensions~\cite{fourd}.  In the latter case, one replaced the surplus
dimensions by
    internal world-sheet degrees of freedom.  Such models could accommodate a
    realistic GUT-like or Standard-Model-like gauge group in four
dimensions~\cite{AEHN}.
    
    Perturbative string models had a number of phenomenological successes. 
    They predicted that there should be at most ten dimensions, in agreement
    with the observed number of four.  They predicted that the gauge group
    should have rank no greater than 22, in agreement with the rank four 
of the
    Standard Model.  They could not accommodate matter in large representations
    of the Standard Model gauge group, such as the \underline 3 of SU(2) or the
    \underline 8 of SU(3), also in agreement with experiment.
Also, in a
    generic string model the maximum size of a Yukawa coupling $\lambda$, such
    as that of the top quark, is of the same order as the gauge coupling
$g$~\cite{Witten}. 
    One therefore predicts a maximum quark mass $m_t = {\cal O}(g)\langle H \rangle$,
    which yields the successful prediction $m_t \lappeq$190~GeV after including
    renormalization-group effects.
    
    Perturbative string theories also permit a calculation of the gauge
    unification scale from first principles~\cite{topdown}:
    \beq
    m_{\rm GUT} = 0(g) \times \frac{m_P}{\sqrt{8\pi}} \simeq {\rm few} \times
    10^{17}~{\rm GeV}~.
    \label{eleven}
    \eeq
    This top-down calculation is close to the bottom-up estimate of $m_{\rm
    GUT}$ based on the experimental values of the gauge couplings:  $m_{\rm
    GUT} \sim 2 \times 10^{16}$~GeV~\cite{bottomup}. but there is an
apparent discrepancy by
    over an order of magnitude.  This provided motivation for considering
    strongly-coupled string models with an extra (relatively) large spatial
    dimension in the context of $M$ theory~\cite{HW}, as we discuss in the
next section.

\section*{Beyond Strings}
    
    The second string revolution was triggered by the understanding of
    non-perturbative effects in string theory~\cite{Polchinski}, and the
role of
    higher-dimensional extended objects such as membranes,
which do not appear in perturbative string theory.  The lowest
    two-dimensional membrane describes a three-dimensional world volume as it
    propagates, and there are analogous higher-dimensional extended structures.
     These may be obtained as soliton solutions of string theory, with masses
     $m \sim 1/g_s$.  Therefore, they become light in the strong-coupling
     limit, just as magnetic monopoles become light in the strong-coupling
     limit of gauge theory.  It has been known for some time that there is a
     duality in gauge theory between these monopoles and conventional gauge
     particles~\cite{MO}.  Now it is known that similar dualities exist in
string theory:
      solitons in some strongly-coupled theories can be identified with light
      states in the weakly-coupled perturbative limit of some other string
      theory~\cite{duality}.  A related development has been the
realization that the string
      coupling $g_s$ should be interpreted as the expectation value of a
      quantum field, that can be interpreted as an extra dimension.  In the
      strong-coupling limit $g_s \rightarrow \infty$, this extra dimension also
      becomes large:  $R \rightarrow \infty$, and a ten-dimensional
      perturbative string theory becomes a non-perturbative eleven-dimensional
      theory~\cite{HW}.  It is now believed that all string theories are
related by
      dualities, and that they all may be regarded as particular limits of this
      underlying eleven-dimensional theory, known as $M$ theory.  In the
      low-energy limit, this theory reduces to eleven-dimensional
supergravity~\cite{elevensugra}, but its full spectrum is infinite.
      
      The question now arises:  which limit of $M$ theory best describes the
      physics we know?  An important clue is provided by the bottom-up
calculation of the unification scale using the low-energy
      gauge-coupling measurements at LEP and elsewhere~\cite{bottomup}.
This can be reconciled with the top-down string
calculation~\cite{topdown} if the
      eleventh dimension is relatively large:  $L_{11} \gg 1/m_{\rm
      GUT} \sim 1/({10^{16}}~{\rm GeV}) \gg 1/m_P$.  The idea is
      that the gravitational interactions ``feel" this extra dimension, and so
      become stronger at shorter distances:  the Newton potential $V_N \propto
      1/r \rightarrow 1/r^2$ in five dimensions.  In this way, the
      gravitational coupling may rise more quickly at energies
$E \gappeq 1/L_{11}$, so as to be unified with the gauge interactions
      (that do not feel the extra dimension) as shown in Fig.~3.  The
proposal
      is that the gauge interactions act only on ten-dimensional subspaces of
      the full eleven dimensions, that may be regarded as ``capacitor plates"
      separated by the large eleventh dimension in which gravity acts, as seen
      in Fig.~4~\cite{HW}.  Six dimensions are compactified, as in
weakly-coupled string
      theory, with a size $R \sim 1/m_{\rm GUT}$.  Thus physics is described by
      a five-dimensional effective supergravity theory at energies $L^{-1}_{11}
      < E < m_{\rm GUT}$~\cite{fivedsugra}.



\begin{figure}
\hspace*{30pt}
\centerline{\epsfig{file=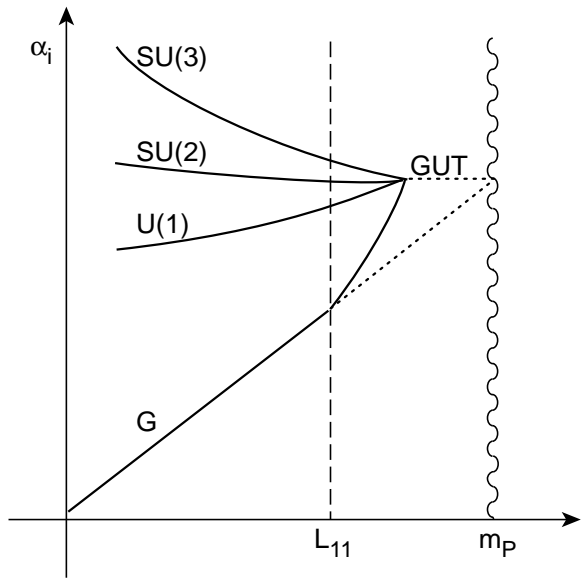,height=3in,width=3in}}
\vspace*{0.5cm}
\caption[]{{\it Sketch of the possible evolution of the gauge couplings
$\alpha_i$ and
the gravitational coupling $G$: if there is a large fifth dimension with
size $\gg m^{-1}_{GUT}$, $G$ may be unified with the gauge couplings at
the GUT scale.}}
\end{figure}

\begin{figure}
\hspace*{30pt}
\centerline{\epsfig{file=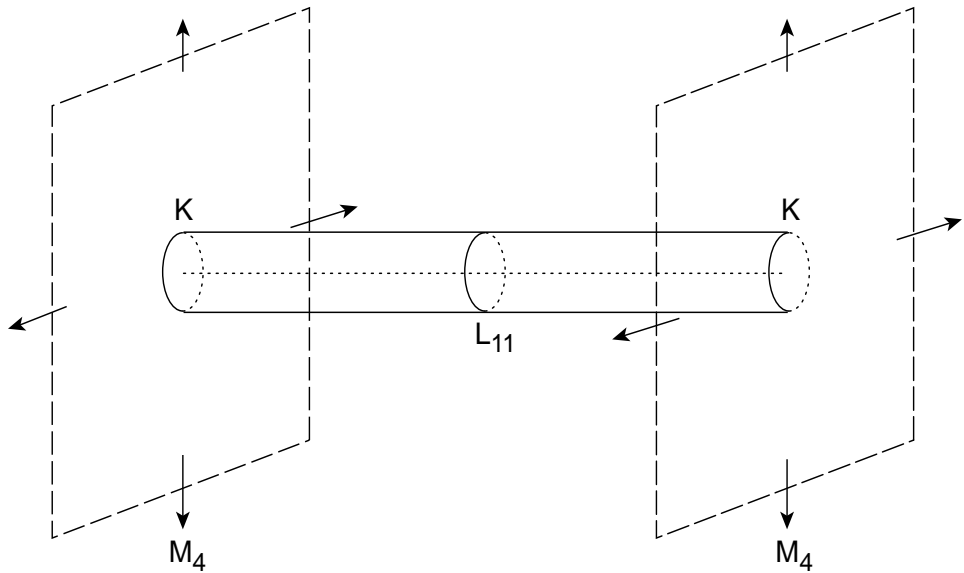,height=3in,width=5in}}
\vspace*{0.5cm}
\caption[]{{\it The capacitor-plate scenario favoured in
eleven-dimensional $M$ theory. The eleventh dimension has a size
$L_{11} \gg M_{GUT}^{-1}$, whereas dimensions $5, ... , 10$ are
compactified on a small manifold $K$ with characteristic size
$\sim M_{GUT}^{-1}$. The remaining four dimensions form
(approximately) a flat Minkowski space $M_4$.}}
\end{figure}

    \section*{An Extra Dimension for Gauge Theories?}
    
    We saw in the previous section how a ten-dimensional string theory becomes
    effectively an eleven-dimensional theory in the strong-coupling limit. 
    Does something similar also happen in gauge theories~\cite{Maldacena}?
Such a
possibility would be
    consistent with the holographic principle~\cite{holography}, according
to which the
    information in a field theory in some space-time can be encoded in another
    field theory on its boundary.  This was proposed originally in the context
    of quantum gravity, as a way of understanding the fact that the black-hole
    entropy is proportional to the area of its event horizon.  In other words,
    all the information in the bulk of a black hole is encoded on its surface.
In the present context, information about the quantum behaviour of a
gauge theory is conjectured to be encoded in the classical behaviour
of a higher-dimensional gravity theory~\cite{Maldacena,jlP}.
    
    The crucial step is to recall that Minkowski space in $d$
dimensions $M_d$ is
    itself the boundary of anti-de-Sitter (AdS) space in $d+1$ dimensions
    (AdS$_{\rm d+1}$), which is described by the metric
    \beq
    ds^2 = \left( 1 + \frac{r^2}{b^2} \right)dt^2 +
    \frac{1}{(1+\frac{r^2}{b^2})}dr^2 + r^2d\Omega^2~,
    \label{twelve}
    \eeq
    where $b$ is the AdS radius.  Is gauge theory on $M_d$ equivalent to
    some theory on AdS$_{d+1}$?  This bulk theory is likely to be a gravity
    theory, because of the Kaluza-Klein trick:
    \beq
    G_{MN}(M,N = 0,1,2,3,4) \rightarrow g_{\mu\nu}, g_{\mu 4}, g_{4\nu}, g_{44}~,
    \label{thirteen}
    \eeq
    according to which the $(\mu 4, 4\nu)$ components of $g_{MN}$ in five
    dimensions are identified as four-dimensional gauge fields.
    
    Thus we arrive at the Maldacena conjecture~\cite{Maldacena}: that a
$d$-dimensional
    conformal (i.e., finite, with no divergences) supersymmetric gauge theory
    may be equivalent, at least in a certain limit, to a conformal supergravity
    (or string) theory in $(d+1)$-dimensional AdS space. The appropriate limit
    in which the conjecture was first formulated was that of large $N_c$ for
    the gauge theory, where the AdS$_{\rm d+1}$ curvature becomes small and
    reliable classical solutions to the supergravity field equations can be
    found.  The spectrum of supergravity excitations in ADS$_{\rm d+1}$ was found
    to correspond to the spectrum of an ${\cal N} = 4$ supersymmetric gauge
    theory in the large-$N_c$ limit, which is known to be finite.  It was also
    suggested that a similar analysis could be made for non-conformal field
    theories.
    
    Witten extended this idea to the ${\cal N} = 4$ supersymmetric theory at
    finite temperature $t \not= 0$~\cite{WittenT}, and was able to
demonstrate magnetic
    confinement as discussed in Section ~3, with the appearance of a mass gap. 
Since the boundary conditions used to study finite-temperature
field theory violate supersymmetry, this provides a prototype
for studying four-dimensional non-supersymmetric
    gauge theories at large $N_c$.  However, it was not possible
in~\cite{WittenT} to demonstrate
asymptotic freedom, and hence that  a different phase did not appear in
the weak-coupling limit.

\section*{A Non-Conformal(ist) Approach}
    
    Conformal symmetry is very powerful:  with no infinities and hence no need
    to introduce a renormalization scale, one can derive many rigorous results.
     However, the real world is  not described by a conformal gauge theory: 
     there is asymptotic freedom, there are quark and gluon
condensates, etc.. 
     So how can one extend the above ideas to go beyond conformal
symmetry~\cite{PolL,EM}?
     
     In string theory, the conformal factor $\phi$ in the world-sheet metric
     \beq
     \gamma_{ab} = e^\phi \hat\gamma_{ab} ,
     \label{fourteen}
     \eeq
     where $\hat \gamma_{ab}$ is a fiducial reference metric, is called a
Liouville field~\cite{DDK}. Perturbatively it
     decouples in a conformal string theory such as the bosonic string in 26
     dimensions or the superstring in ten dimensions, but has non-trivial
     dynamics in a non-conformal string theory, as would be needed to describe
     QCD, which lives in fewer dimensions.  We have suggested that one
     identifies the Liouville field with a $q$-number scale factor, that can be
     identified as a dynamical renormalization scale in non-critical string
     theory~\cite{Liouville}.  We have also discussed the interpretation
of
world-sheet defects~\cite{defects}
     in such a non-critical Liouville string theory as black holes in target
     space~\cite{EMND}.
     
     These ideas have recently been applied to the problem of confinement in
     gauge theory~\cite{EM}.  We have shown that a world-sheet vortex
defect corresponds
     to the intersection of a target-space magnetic-field line with the world
     sheet, as seen in Fig.~5.  This is because one may rewrite the
line integral in (\ref{two}) in the following way:
     \beq
     \int_CA \cdot dl = \int_{\Sigma(C)} B \cdot ds~,
     \label{fifteen}
     \eeq
     where $dS$ is an element of the surface $\Sigma(c)$ that is bounded by the
     curve $C$ whose line element is $dl$.  One may then decompose
     \beq
     B = \epsilon_{ab} \partial_a A_b = \epsilon_{ab} \partial_a \partial_b
     X^\mu A_\mu(x) + ...~,
     \label{sixteen}
     \eeq
     where $\partial_a \partial_bX^\mu \not= 0$ in the presence of a
     world-sheet vortex.  Condensation of magnetic monopoles in target space
     creates a flux through the world sheet that corresponds to condensation of
     vortices on the world sheet.  This condensation is known to occur at a low
     enough temperature $T < T_{\rm vortex}$, and we have shown that vortex
     condensation generates a tension for the string:
     \beq
     \mu = (\alpha')^{-1} \propto \langle(\epsilon_{ab} \partial_a \partial_b X)^2
     \rangle \not= 0
     \label{seventeen}
     \eeq
     corresponding to confinement in the string picture~\cite{EM}.


\begin{figure}[htb]
\hspace*{10pt}
\centerline{\epsfig{file=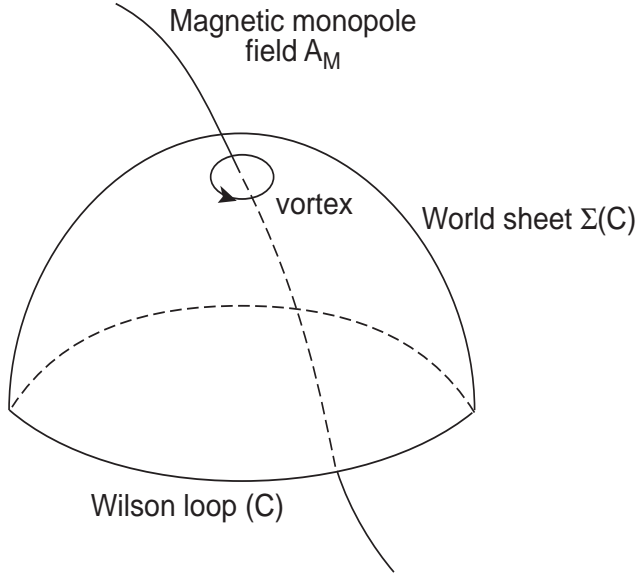,height=3in,width=3.3in}}
\vspace*{0.5cm}
\caption[]{{\it 
A string world sheet $\Sigma(C)$, whose boundary is a Wilson loop $C$,
in the presence of a four-dimensional space-time magnetic field,
as could be generated by a target-space monopole.
The intersection of the field line with the string world sheet
results in a world sheet vortex, which is related by world-sheet duality
to a world-sheet spike. Condensation of magnetic flux in space time
corresponds to condensation of vortices on the world sheet,
which generates a string tension~\cite{EM}.}} 
\label{fig5}
\end{figure}

     We have also shown~\cite{EM} how these ideas can be extended to AdS
black holes, as illustrated in Fig.~6. Black holes
are related to defects of different type, called spikes, that are
     related to world-sheet vortex defects by a world-sheet
duality~\cite{defects}. Stable
     AdS black holes correspond to a condensation of spikes on the world sheet,
     which is known to occur at a high enough temperature $T > T_{\rm spike} >
     T_{\rm vortex}$.  The AdS gravity theory has phases~\cite{HP} that
parallel those of QCD:
     \bea
     &T < T_{\rm vortex}:&  {\rm
     vortices~condense,~no~black~holes,~confinement}\nonumber \\
     &T_{\rm vortex} < T < T_{\rm spike}:&  
{\rm vortices~unbound,~black~holes~unstable,~mixed~phase}\nonumber \\
&T_{\rm spike} < T:& {\rm spikes~condense, stable~black~
     holes,~deconfinement}
     \eea
     The temperature dependence of the free energy resembles that shown in
     Fig.~1 for QCD, with $T_2 \leftrightarrow T_{\rm spike}$ and $T_0
     \leftrightarrow T_{\rm vortex}$.  The behaviour of the bulk theory is
     reminiscent of a liquid-gas transition~\cite{EGM}, a possibility to
which we return shortly.


\begin{figure}[htb]
\hspace*{10pt}
\centerline{\epsfig{file=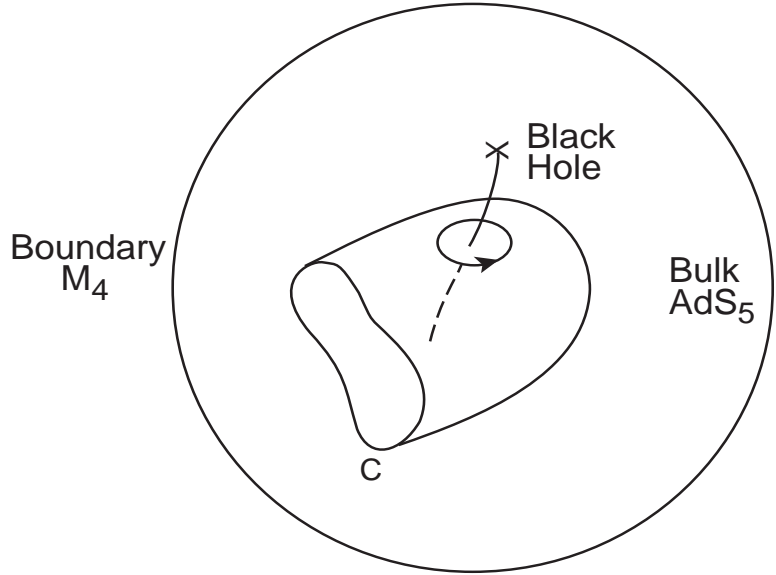,height=3in,width=4in}}
\vspace*{0.5cm}
\caption[]{{\it
The string world sheet of Fig.~5 is embedded in four-dimensional
Minkowski space M$_4$, which is regarded here as the boundary
of five-dimensional Anti-de-Sitter space AdS$_5$. A black hole in
the bulk of AdS$_5$ corresponds to a vortex on the world sheet,
whose condensation corresponds to confinement~\cite{EM}.}}
\label{fig6}
\end{figure} 

\section*{What does this have to do with Heavy Ions?}
     
     These advanced new theoretical ideas certainly provide deeper
     understanding and insight into confinement and deconfinement in
QCD~\cite{jlP}. They
     may also provide useful new calculational techniques.  Concretely, one may
     hope to tackle non-perturbative strong-coupling phenomena in
     four-dimensional QCD using the techniques of classical weakly-coupled
     gravity or string theory in AdS$_5$.  Indeed, people have started to
     calculate quantities of experimental interest.
     
     {\it Glueball Masses~\cite{balls}:}  correlation functions in QCD$_4$
in the limit of
     large $N_c$ can be related to classical properties of supergravity in
     AdS$_5$.  Particle poles can therefore be related to classical eigenvalue
     equations.  In this way, calculations in a certain limit have been made of the ratios of
     glueball masses, e.g.,
     \beq
     \frac{m_{0^{-+}}}{m_{0^{++}}} = 1.20~,
     \label{nineteen}
     \eeq
     to be compared with the ratio $1.36 \pm 0.32$ obtained from a lattice
     computation.  It remains to be seen how results in the AdS$_5$ supergravity limit
     correspond to the physical case of QCD$_4$, but at least one has some numbers
     to discuss.
     
     {\it Vacuum Properties~\cite{vacuum}}:  First calculations have been
made of the gluon
     condensate and topological susceptibility $\chi_t$, with the results:
     \beq
     \frac{\langle 0|GG|0 \rangle}{\langle 0|GG|0\rangle \vert_{\rm lattice}} \simeq
     0.9~,~~~ \frac{\chi_t}{\chi_t \vert_{\rm lattice}} \simeq 1.7~,
     \label{twenty}
     \eeq
     as well as the string tension:
     \beq
     \frac{\mu}{\mu|_{\rm lattice}} \simeq 0.4~.
     \label{twentyone}
     \eeq
     These numbers are not perfect, and the AdS$_5$ supergravity limit does not
     correspond to the physical QCD$_4$ limit, but please remember that these are early days
  yet.
  
  {\it Heavy-Quark Potential~\cite{potential}}:  First calculations are
quite promising, with
  the expected form:
  \beq
  V(R) \sim \mu R + {\rm (cont.)} + {\rm exp}(-R \Lambda_{QCD}) + \dots~,
  \label{twentytwo}
  \eeq
  emerging.
  
  Finally, what about heavy ions?  We are trying to model the behaviour of the
  free energy and pressure near the QCD transition~\cite{EGM}, which are
known from
  lattice calculations to exhibit departures from an ideal gas.    This effect
  is particularly marked for the pressure, which is considerably less than the
  ideal-gas value, as seen in Fig.~7~\cite{lattice}.  One interpretation
of this departure has
  been in terms of an effective magnetic mass for the gluon above the
  transition temperature.
  

\begin{figure}[htb]
\vspace*{0.5cm}
\hspace*{30pt}
\centerline{\epsfig{file=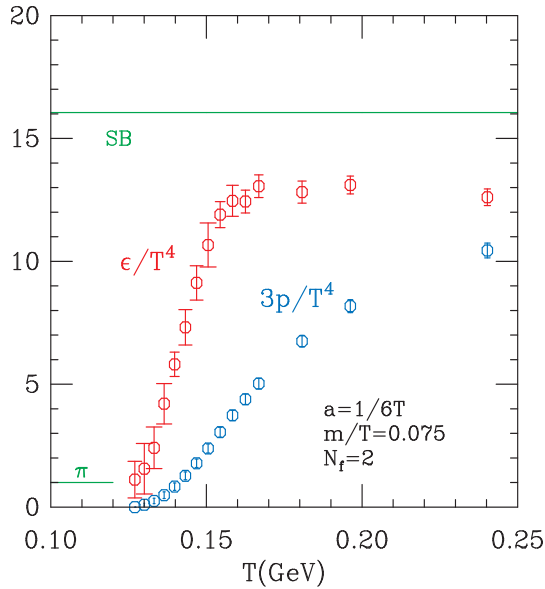,height=3in,width=3in,}}
\vspace*{0.5cm}
\caption[]{{\it Scaled energy density $\epsilon/T^4$ compared with
three times the scaled pressure $3p/T^4$, as obtained from
lattice QCD calculations: for a discussion and original
references, see~\cite{lattice}.}}
\label{fig7}
\end{figure}

  We regard this behaviour as a manifestation of a  liquid-gas transition for
  black holes in AdS$_5$, which are regarded as effective
regulators for the monopoles in $M_4$.  These black
holes interact via a non-trivial
  effective potential with analogues of van der Waals forces.  A standard
  analysis then leads to a van der Waals formula for the equation of state,
  \beq
  RT = (V-b) \left( P + \frac{a}{v^2}\right)~,
  \label{twentythree}
  \eeq
  leading to the familiar pattern of isothermal pressure curves.  
Based on this picture, we obtain the behaviours of the free energy
  and pressure shown in Fig.~8 near the critical temperature
$T_c$~\cite{EGM}. These have
  the right qualitative behaviour, including the vanishing of the pressure close to
  $T_c$.
  

\begin{figure}[htb]
\vspace*{30pt}
\centerline{\epsfig{file=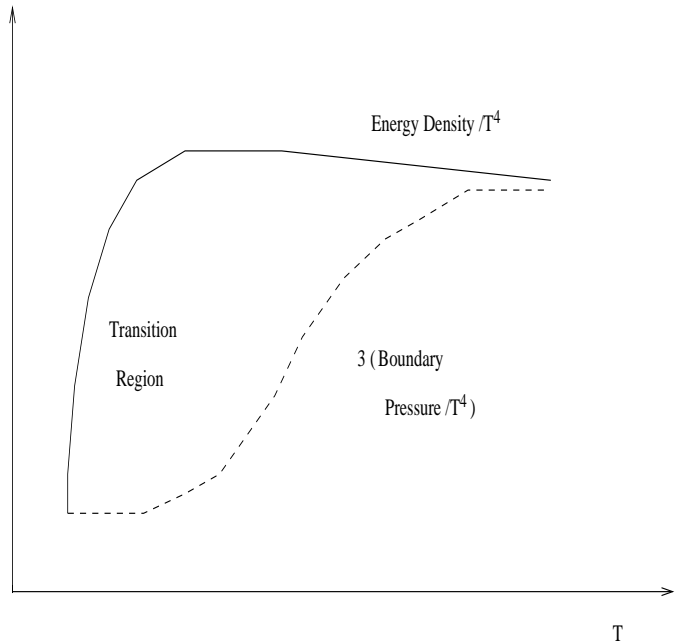,height=3.5in,width=3.5in}}
\vspace*{0.5cm}
\caption[]{{\it
Scaled energy density and
three times the scaled pressure obtained from 
a preliminary analysis using an
effective van der Waals description of a non-ideal gas of
AdS$_5$ black holes~\cite{EGM}. The
qualitative behaviour, in particular the fact that the scaled
energy density is larger than $3p/T^4$ in the transition region, is 
similar to that in the lattice calculations shown in Fig.~7.}}
\label{threepre}
\end{figure}

   We conclude on the optimistic note that modern
ideas~\cite{Polchinski,Maldacena,EM} in non-perturbative
   string theory duality and conformal field theory may indeed have some
   relevance to heavy-ion physics.  Some non-perturbative QCD quantities have
   already been estimated~\cite{balls,vacuum,potential}, and more are on
the way. Not only qualitative
   insights but also quantitative results may eventually emerge.  The
   finite-temperature QCD transition may be yielding to the theoretical
   assault~\cite{EGM}.  Perhaps these ideas may even eventually enable the
calculation of
   physical quantities measurable directly in heavy-ion collisions?


\begin{references}
\bibitem{chilag} J. Gasser and H. Leutwyler, {\it Phys. Rep.} 87 (1987) 
77.

\bibitem{intout} A. Pich, hep-ph/9806303.

\bibitem{Skyrme} E. Witten {\it Nucl.Phys.} B223 (1983) 422, 433;\\
G.S. Adkins, C. Nappi and E. Witten, {\it Nucl.Phys.} B228 (1983) 552.

\bibitem{BEK} S.J. Brodsky, J. Ellis and M. Karliner, {\it Phys.Lett.}
B206 (1988) 309.

\bibitem{nspin} Spin Muon Collaboration, B. Adeva et al., {\it Phys.Rev.}
D58 (1998) 112002.

\bibitem{BG} P. Binetruy and M.K. Gaillard, {\it Phys.Rev.} D32 (1985)
931.

\bibitem{GL} J. Gasser and H. Leutwyler,
{\it Phys.Lett.} 184B (1987) 83 and 188B (1987) 477.

\bibitem{CEO} B.A. Campbell, J. Ellis and K.A. Olive,
{\it Phys.Lett.} B235 (1990) 325 and {\it Nucl.Phys.} B345 (1990) 57.

\bibitem{EGM} J. Ellis, A. Ghosh and N.E. Mavromatos, CERN preprint
TH/99-43 in preparation.

\bibitem{EG} J. Ellis and K. Geiger, {\it Nucl.Phys.} A590 (1995) 609c and
{\it Phys.Rev.} D52 (1995) 1500.

\bibitem{VNI} K. Geiger, {\it Comput.Phys.Commun.} 104 (1997) 70.

\bibitem{coffin} J. Gasser and H. Leutwyler, {\it Nucl.Phys.} B307 (1988)
763.

\bibitem{EGK} J. Ellis, K. Geiger and H. Kowalski,
{\it Phys.Rev.} D54 (1996) 5443.

\bibitem{EGWW} J. Ellis and K. Geiger, {\it Phys.Rev.} D54 (1996) 1967 and
{\it Phys.Lett.} B404 (1997) 230.

\bibitem{GEHW} K. Geiger, J. Ellis, U. Heinz and U. Wiedemann,
hep-ph/9811270.

\bibitem{UHUW} U. Heinz and U. Wiedemann, talks at this meeting.

\bibitem{SW} N. Seiberg and E. Witten, {\it Nucl.Phys.} B431 (1994) 484.

\bibitem{monopoles} S. Mandelstam, {\it Phys.Rep.} 23C (1976) 245;\\
G. `t Hooft, {\it Nucl.Phys.} B190 [FS3] (1981) 455.

\bibitem{Pol}  A.M. Polyakov, {\it Nucl.Phys.} B72 (1974) 461;
{\it Phys.Lett.} B59 (1975) 79;
{\it ibid.} B268 (1986) 406;
{\it Gauge Fields and Strings} (Harwood, 1987);
hep-th/9711002.

\bibitem{Veneziano} G. Veneziano, {\it Nuovo Cim.} A57 (1968) 190.

\bibitem{string} P. Goddard, J. Goldstone, C. Rebbi and C. Thorne,
{\it Nucl.Phys.} B56 (1973) 109.

\bibitem{SS} J. Scherk and J.H. Schwarz, \NP B81 (1974) 118.

\bibitem{Polchinski} J. Polchinski, {\it Phys.Rev.Lett.} 75 (1995) 184;
\\
J. Polchinski, S. Chaudhuri and C. Johnson, hep-th/9602052
and references therein; \\
J. Polchinski, TASI lectures on D-branes, hep-th/9611050, and references
therein.

\bibitem{Lovelace} C. Lovelace, {\it Phys.Lett.} 34B (1971) 500.

\bibitem{GS} M.B. Green and J.H. Schwarz, {\it Phys.Lett.} 149B
(1984) 117 and {\it Phys.Lett.} 151B (1985) 21.

\bibitem{heterotic} D.J. Gross, J.A. Harvey, E.
Martinec and R. Rohm, {\it Phys.Rev.Lett.} 54 (1985) 502.

\bibitem{quartet}  P. Candelas, G. T. Horowitz, A. Strominger and
E. Witten, {\it Nucl.Phys.} B258 (1985) 46.

\bibitem{fourd} See, e.g., I. Antoniadis, C. Bachas and C. Kounnas,
{\it Nucl.Phys.} B289 (1988) 87;\\
I. Antoniadis and C. Bachas, {\it Nucl.Phys.} B298 (1987) 586.

\bibitem{AEHN} See, e.g., I. Antoniadis, J. Ellis, J.S. Hagelin and D.V.
Nanopoulos, {\it Phys.Lett.} B231 (1989) 65.

\bibitem{Witten} E. Witten, {\it Phys.Lett.} B155 (1985) 151.

\bibitem{topdown} See, e.g., I. Antoniadis, J. Ellis, R. Lacaze and D.V.
Nanopoulos, {\it Phys.Lett.} B268 (1991) 188.

\bibitem{bottomup} J. Ellis, S. Kelley and D.V. Nanopoulos, {\it
Phys.Lett.} B249
(1990) 441 and {\it Phys.Lett.} B260 (1991) 131;\\
U. Amaldi, W. de Boer and H. Furstenau, {\it Phys.Lett.} B260 (1991)
447;\\
P. Langacker and M. Luo, {\it Phys.Rev.}  D44 (1991) 817.

\bibitem{HW} P. Horava and E. Witten, {\it Nucl.Phys.} {B460} (1996)
506;\\
E. Witten, {\it Nucl.Phys.} {B471} (1996) 135;\\
P. Horava and E. Witten, {\it Nucl.Phys.} {B475} (1996) 94;\\
T. Banks and M. Dine, {\it Nucl.Phys.} {B479} (1996) 173.

\bibitem{MO} C. Montonen and D.I. Olive, {\it Phys.Lett.} 72B (1997) 117.

\bibitem{duality} For a review, see: A. Sen, hep-ph/9810356.

\bibitem{elevensugra} E. Cremmer, B. Julia and J. Scherk,
{\it Phys.Lett.} 76B (1978) 409;\\
E. Cremmer and B. Julia, {\it Phys.Lett.} 80B (1978) 48 and {\it
Nucl.Phys.} B159 (1979) 141.

\bibitem{fivedsugra} I. Antoniadis, S. Ferrara and T. Taylor,
{\it Nucl.Phys.} B460 (1996) 489;\\
A. Lukas, B.A. Ovrut and D. Waldram, hep-th/9803235;\\
A. Lukas, B.A. Ovrut, K.S. Stelle and D. Waldram, hep-th/9806051;\\
R. Donagi, A. Lukas, B.A. Ovrut and D. Waldram, hep-th/9811168;\\
J. Ellis, Z. Lalak, S. Pokorski and W. Pokorski,
{\it Nucl.Phys.} B540 (1999) 149;\\
J. Ellis, Z. Lalak and W. Pokorski, hep-th/9811133.

\bibitem{Maldacena} J. Maldacena, {\it Adv.Theor.Math.Phys.} 2 (1998) 231.

\bibitem{holography} G. `t Hooft, gr-qc/9310026; \\
L. Susskind, hep-th/9409089.

\bibitem{jlP} For a theoretical review of this explosive field, see:\\
J.L. Petersen, hep-th/9902131.

\bibitem{WittenT} E. Witten, {\it Adv.Theor.Math.Phys.} 2 (1998) 505.

\bibitem{PolL} S. Gubser, I. Klebanov and A.M. Polyakov, hep-th/9802109.

\bibitem{EM} J. Ellis and N.E. Mavromatos, CERN-TH-98-264,
hep-th/9808172.

\bibitem{DDK} F. David, {\it Mod.Phys.Lett.} A3 (1988) 1651;\\
J. Distler and H. Kawai, {\it Nucl.Phys.} B321 (1989) 509.

\bibitem{Liouville} J. Ellis, N.E. Mavromatos
and D.V. Nanopoulos, {\it Phys.Lett.} B293 (1992) 37;
{\it Mod.Phys.Lett.} A10 (1995) 425; \\
Lectures presented at the  
{\it Erice Summer School, 31st Course: From Supersymmetry to the
Origin of Space-Time},
Ettore Majorana Centre, Erice, July 4-12
1993 ; hep-th/9403133, `Subnuclear Series' Vol. 31,
(World Scientific, Singapore 1994), p.1

\bibitem{defects} 
J. Cardy, {\it Nucl.Phys.} B205 (1982) 17; \\
J. Cardy and E. Rabinovici, {\it Nucl.Phys.} B205 (1982), 1;\\
B. Ovrut and S. Thomas, {\it Phys.Lett.} B237 (1991)
292;\\
J. Ellis, N.E. Mavromatos and D.V. Nanopoulos,
{\it Phys.Lett.} B289 (1992) 25 and
{\it Mod.Phys.Lett.} A12 (1997) 2813.

\bibitem{EMND} J. Ellis, N.E. Mavromatos and D.V. Nanopoulos,
{\it Int.J.Mod.Phys.} A12 (1997) 2639.

\bibitem{HP}  S. Hawking and D. Page, {\it Comm.Math.Phys.} 87 (1983)
577.

\bibitem{balls} C. Cs\'aki,
H. Ooguri, Y. Oz and J. Terning, preprint UCB-PTH-98-30,
hep-th/9806021; \\
R. de Mello Koch, A. Jevicki, M. Mihailescu and J. P.
Nunes, preprint BROWN-HET-1130, hep-th/9806125; \\
M. Zyskin, \PL B439 (1998) 373; \\
H. Ooguri, H. Robins and J. Tannenhauser, \PL B437 (1998) 77;\\
J. Minahan, hep-th/9811156.

\bibitem{vacuum} A. Hashimoto and Y. Oz, hep-th/9809106.

\bibitem{potential} J. Greensite and P. Olesen, {\it JHEP} 9808 (1998)
009 and hep-th/9901057.

\bibitem{lattice} For a recent review, see: H. Satz, 
hep-ph/9711289.

\end{references}
\end{document}